# $^{177}$Lu SPECT Imaging in the Presence of $^{90}$Y: Does $^{90}$Y Degrade Image Quantification? A Simulation Study

**Cassandra Miller[1,2], Carlos Uribe[1,3,4], Xinchi Hou[1,4], Arman Rahmim[1,2,4] and Anna Celler[4]**

[1]Department of Integrative Oncology, BC Cancer Research Institute, Vancouver, BC, Canada
[2]Department of Physics, University of British Columbia, Vancouver, BC, Canada
[3]Functional Imaging, BC Cancer, Vancouver, BC, Canada
[4]Department of Radiology, University of British Columbia, Vancouver, BC, Canada

E-mail: cassandramiller@phas.ubc.ca, curibe@bccrc.ca, arman.rahmim@ubc.ca

**Abstract**

This work aims to investigate the accuracy of quantitative SPECT imaging of $^{177}$Lu in the presence of $^{90}$Y, which occurs in dual-isotope radiopharmaceutical therapy (RPT) involving both isotopes. We used the GATE Monte Carlo simulation toolkit to conduct a phantom study, simulating spheres filled with $^{177}$Lu and $^{90}$Y placed in a cylindrical water phantom that was also filled with activity of both radionuclides. We simulated multiple phantom configurations and activity combinations by varying the location of the spheres, the concentrations of $^{177}$Lu and $^{90}$Y in the spheres, and the amount of background activity. We investigated two different scatter window widths to be used with triple energy window (TEW) scatter correction. We also created multiple realizations of each configuration to improve our assessment, leading to a total of 540 simulations. Each configuration was imaged using a simulated Siemens SPECT camera. The projections were reconstructed using the standard 3D OSEM algorithm, and errors associated with $^{177}$Lu activity quantification and contrast-to-noise ratios (CNRs) were determined. In all configurations, the quantification error was within ± 6% of the no-$^{90}$Y case, and we found that quantitative accuracy may slightly improve when $^{90}$Y is present because of reduction of errors associated with TEW scatter correction. The CNRs were not significantly impacted by the presence of $^{90}$Y, but they were increased when a wider scatter window width was used for TEW scatter correction. The width of the scatter windows made a small but statistically significant difference of 1-2% on the recovered $^{177}$Lu activity. Based on these results, we can conclude that activity quantification of $^{177}$Lu and lesion detectability is not degraded by the presence of $^{90}$Y.

## 1. Introduction

Radiopharmaceutical therapy (RPT) is a safe and increasingly promising therapy for various types of cancers, such as neuroendocrine tumours (NETs) and prostate cancer [1]. In these therapies, a pharmaceutical is labelled with a radioisotope and injected into the patient where it binds to receptors on tumour cells. The proximity of the radioisotope to the tumour allows the particulate emissions to damage the DNA of the tumour cells as the radioisotope decays. If the radionuclide is also a gamma emitter, Single Photon Emission Computed Tomography (SPECT) images can be obtained during therapy to perform dosimetry calculations to determine radiation doses delivered to tumours and organs at risk [1].

Two commonly used radionuclides for RPT are $^{177}$Lu and $^{90}$Y, which have been used to label different pharmaceuticals such as DOTATATE/DOTATOC and PSMA-617 [2–4]. $^{90}$Y is also commonly used for radioembolization of liver tumours [5]. While typically only one radionuclide at a time is used for RPT, studies have shown that combination (or 'tandem') RPTs, with simultaneous administration of two radionuclides, e.g. $^{177}$Lu/$^{90}$Y-DOTATATE for NETs or $^{177}$Lu/$^{225}$Ac-PSMA-617 for prostate cancer [6–8] may have favourable outcomes compared to using only one radionuclide. Presently, tandem therapy with $^{177}$Lu and $^{90}$Y is mainly pursued for peptide receptor radionuclide therapy (PRRT) for NETs and may expand to other cancer types in the future.

A pre-clinical study by de Jong et al. [9] showed that for a combination of $^{177}$Lu and $^{90}$Y-labelled analogues, the median survival time was at least doubled compared to using $^{90}$Y or $^{177}$Lu alone after PRRT treatment. Furthermore, Kunikowska et al. [10] demonstrated in a clinical study that such combination therapy resulted in longer survival times compared to using $^{90}$Y alone with a similar safety risk (also for NETs). Patients treated with $^{90}$Y only reached a survival time of 26.2 months, while the survival time was not reached after 48 months for those receiving the combination therapy.

RPT for most cancers is standardized, with treatment typically being administered over 4-6 cycles, and most patients receiving an intravenous injection per cycle of 7.4 GBq for $^{177}$Lu [11] and 3.7 GBq/m$^2$ for $^{90}$Y [12]. Unfortunately, this lack of personalized treatment may result in patients being treated sub optimally; current treatment protocols result in tumour remission rates of only 15-35% [13,14]. These outcomes may be partially due to a variation in absorbed tumour doses among patients and the absence of personalized treatment protocols in clinics. Several studies have demonstrated that radiotracer uptake in tumours and critical organs may vary widely between patients [15,16]. Treatment outcomes may improve if patient specific treatment protocols, based on personalized dosimetry, were implemented to determine the optimal and safe activities to be administered for each patient.

Internal dosimetry calculations require quantitative information about the activity distribution in the patient's body and its biodistribution. This information can be obtained by acquiring quantitative images at multiple time points, from which the time-integrated activity, required for dose estimates, can be calculated. Using this approach, Del Prete et al. [17] showed that in some cases the injected activity of $^{177}$Lu could be increased to 10.9 ± 5.0 GBq per cycle while still being under the dose limit for organs at risk such as the kidneys. A fixed injected activity of 7.4 GBq may not be optimal for these patients.

The decay of $^{177}$Lu results in emissions of both betas and gammas, making it useful for treatment while also allowing personalized dosimetry calculations using data obtained from quantitative SPECT images. Meanwhile, dosimetry cannot as easily be performed for $^{90}$Y because it is a pure beta emitter. Bremsstrahlung SPECT imaging of $^{90}$Y is sometimes performed, but mainly for visualization of activity distribution, and quantitative Bremsstrahlung measurements are very challenging [18,19] and not routinely performed. Similarly, although Positron Emission Tomography (PET) imaging of $^{90}$Y's extremely weak positron emissions has been proposed, activity quantification in this case is also difficult [20,21]. Commonly, dosimetry of $^{90}$Y is only performed when high activities are injected, such as those used in radioembolization procedures [5,22]

If the biodistributions of the radiopharmaceuticals used for tandem therapy are similar, then the quantitative information about radiopharmaceutical distribution (such as its uptake and washout) derived from $^{177}$Lu imaging studies can be used to determine the dose from $^{90}$Y. It has been shown that the same pharmaceutical labelled with different radionuclides likely has different biodistributions in the body, however the differences may be small enough and the patient benefits of dosimetry great enough that using one radionuclide to determine the patient dose from another radionuclide may be worth it [23].

However, if both isotopes are present in the body at the same time, quantitative imaging of one radionuclide can be challenging due to crosstalk, or the presence of photons from one radionuclide in the energy window of the other radionuclide. In the case of $^{177}$Lu and $^{90}$Y, the Bremsstrahlung photons created by the high energy beta emissions from $^{90}$Y might contaminate the $^{177}$Lu spectrum, resulting in degraded images. This could result in quantitative errors and therefore inaccurate dosimetry estimates, possibly leading to suboptimal treatment of patients. Therefore, before accurate dosimetry can be attempted, the quantitative accuracy of $^{177}$Lu SPECT imaging in the presence of $^{90}$Y must be determined.

The aim of the present work is to determine the accuracy of quantitative imaging, with respect to noise and activity quantification, of $^{177}$Lu in the presence of $^{90}$Y. Multiple simulations of cylindrical phantoms containing spheres filled with various activities of $^{177}$Lu, $^{90}$Y, or both, were created and imaged with a simulated SymbiaT SPECT imaging system. In each case, the accuracy of $^{177}$Lu activity quantification and the contrast to noise ratio (CNR) was evaluated to further assess the impact of the presence $^{90}$Y on image signal and noise.

SPECT image quantification requires performing accurate scatter correction, most often using the triple energy window scatter correction (TEW SC) method [24]. Because the Bremsstrahlung from $^{90}$Y will contribute photons to the entire $^{177}$Lu spectrum, we also evaluated the accuracy of activity quantification and CNR for two different scatter window widths used in the TEW method.

## 2. Methods

*2.1 Monte Carlo Simulations*

Five hundred and forty simulations of spheres filled with $^{177}$Lu, $^{90}$Y, or both isotopes, placed inside of a water cylinder, were performed. The Geant4 Applications for Tomographic Emission (GATE) version 8.0 [25] Monte Carlo code was used for all simulations, and the Geant4 Penelope model was used to model the physics phenomena, such as the scattering, ionization, and Bremsstrahlung. Agreements between our simulation model and phantom measurements were achieved in previous works [26] [27].

The GATE *UserSpectrum* was used to simulate $^{177}$Lu, and the *fastY90* source was used to simulate $^{90}$Y. The *fastY90* source uses a Bremsstrahlung kernel to directly generate Bremsstrahlung photons, instead of simulating the full electron transport of the beta particles [25].

To reduce simulation time, only x-ray and gamma emissions from $^{177}$Lu were simulated. Although beta particles from $^{177}$Lu also create Bremsstrahlung radiation, Uribe et al. [27] showed that $^{177}$Lu Bremsstrahlung photons detected by SPECT never exceed 0.2% of the total photons detected from $^{177}$Lu. Therefore, they were not included in our simulations. The decay characteristics of $^{177}$Lu and $^{90}$Y are provided in table 1.

Table 1 Decay characteristics for $^{177}$Lu and $^{90}$Y [28][29].

| Isotope | Half Life (days) | X-Ray and Gamma Emissions (keV) | Max β Energy (keV) | Mean Range in Tissue (mm) | Max Range in Tissue (mm) |
|---|---|---|---|---|---|
| $^{177}$Lu | 6.64 | 7.9 (3.04%)<br>54.6 (1.57%)<br>55.8 (2.71%)<br>63.9 (0.30%)<br>63.2 (0.59%)<br>64.9 (0.2%)<br>71.6 (0.17%)<br>112.9 (6.17%)<br>136.7 (0.047%)<br>208.4 (10.36%)<br>249.7 (0.20%)<br>321.3 (0.21%) | 496.8 | 0.5 | 2.0 |
| $^{90}$Y | 2.56 | None | 2280.1 | 4.1 | 11.3 |

Three different phantom configurations were simulated (see figure 1):

I. A single sphere (sphere A) filled with $^{177}$Lu placed in a cylindrical phantom filled with a background activity concentrations:
   A. 10 times lower than that in sphere A

   B. 5 times lower than that in sphere A

II. A single sphere (sphere A) filled with both $^{177}$Lu and $^{90}$Y placed in the phantom filled with background activity concentrations:
   A. 10 times lower than that of $^{177}$Lu in sphere A, and 10 times lower than that of $^{90}$Y in sphere A

   B. 5 times lower than that of $^{177}$Lu in sphere A, and 5 times lower than that of $^{90}$Y in sphere A

III. Two spheres, one filled with $^{177}$Lu (sphere A) and the other filled with $^{90}$Y (sphere B) placed in the phantom filled with background activity concentrations:
   A. 10 times lower than that of $^{177}$Lu in sphere A, and 10 times lower than that of $^{90}$Y in sphere B

The spheres were 3.3 cm in radius (150.5 ml) and filled with a uniform solution of $^{177}$Lu, $^{90}$Y, or both isotopes. This sphere size was chosen to replicate the volume of an average kidney, which is around 150 ml [30]. The spheres were placed in the central plane of a cylindrical phantom that had a height of 20.0 cm and a radius of 10.0 cm (6283 ml). In configuration III, when both sphere A and sphere B were present, they were adjacent to each other with the center of each sphere being 3.3 cm from the phantom center as shown in figure 1. The phantom was placed in the center of the field of view of the SPECT camera.

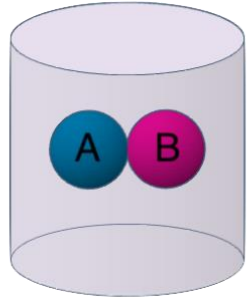

Figure 1 Schematic representation of the 20 x 20 cm cylindrical water phantom containing two 150.5 ml spheres. In configuration I and II, only sphere A was present, filled with $^{177}$Lu and/or $^{90}$Y. In configuration III, sphere A was filled with $^{177}$Lu and sphere B was filled with $^{90}$Y.

Configuration I was investigated to provide a baseline or 'truth' to estimate the quantitative accuracy of imaging different $^{177}$Lu activities, scanned alone, without $^{90}$Y present. Configuration II models the scenario in which both isotopes concentrate in a similar location. Configuration III was used to determine what effect the $^{90}$Y activity located outside the region/organ containing $^{177}$Lu would have on activity quantification. Such effect would be due to either the primary and scattered Bremsstrahlung photons generated by $^{90}$Y (and is dependent on the $^{90}$Y source activity and its location inside the phantom) or to both isotopes having different biodistributions.

No simulations with cold (non-radioactive) water in the cylinder were analysed in this study because cold background is not expected to occur in patient studies. Furthermore, the quantified activity in such configuration has been shown to be overestimated due to attenuation effects and underestimation of scatter in the ROI by the triple energy window scatter correction method [26].

To make this study more realistic while still using a very simple configuration, we based the activity concentrations of $^{177}$Lu in the spheres on patient data obtained from our collaborators at Université Laval in Quebec. Thirty-three patients were injected with 4.1 GBq – 26.2 GBq (average: 9.8 ± 5.0 GBq) of $^{177}$Lu-DOTATATE, which resulted in a kidney concentration ranging from 0.08 MBq/ml to 1.12 MBq/ml (average: 0.56 ± 0.25 MBq/ml). These concentrations did not scale with the injected activity, but instead were patient specific, with the percent of the injected activity absorbed in the kidneys ranging from 0.27% to 1.27% (average: 0.88 ± 0.35%). These values were calculated using data from patients' SPECT scans taken approximately 4 hours post injection. To reflect this data in our simulations, activity concentrations of 0.1 MBq/ml, 0.8 MBq/ml and 1.5 MBq/ml were chosen for sphere A containing $^{177}$Lu to represent patients with low, medium and high uptake. While these numbers are not perfectly representative of what one would find in a true dual-isotope therapy, because uptake is so patient specific it would be challenging to find a one size fits all activity concentration. We have chosen the highest and lowest uptakes to obtain a significantly large range.

In configurations II.A, II.B, and III, five different ratios (0.25:1, 0.5:1, 1:1, 3:1, and 6:1) of $^{90}$Y to $^{177}$Lu activity were simulated in the spheres for each $^{177}$Lu concentration. This was done to assess how different amounts of $^{90}$Y affect $^{177}$Lu activity quantification. All the simulated activities are summarized in table A.1 in the supplementary material.

The $^{177}$Lu to $^{90}$Y activity concentration ratios of 1:3 and 1:6 are unlikely to be observed in any clinical scenario [12][10]. Since $^{90}$Y emits higher energy beta particles than $^{177}$Lu, it is deemed more toxic to organs at risk; injected activities are usually lower. However, these ratios were included in our simulations to determine if the limit at which the error of $^{177}$Lu activity quantification exceeds ±10% of the $^{177}$Lu-only scenario (configuration I) is reached, which we would consider to be the upper limit of acceptable quantification error.

To reduce the simulation time, seven of the simulation components were simulated separately. These seven components are listed in table 2 along with the total number of decays simulated for each one.

Table 2 Total number of decays simulated with GATE for each simulation component.

| Simulation Component | Number of Decays Simulated |
|---|---|
| Sphere A filled with 0.1 MBq/ml of $^{177}$Lu | $1.4 \times 10^{10}$ |
| Sphere A filled with 0.8 MBq/ml of $^{177}$Lu | $1.2 \times 10^{11}$ |
| Sphere A filled with 1.5 MBq/ml of $^{177}$Lu | $2.2 \times 10^{11}$ |
| Sphere A filled with $^{90}$Y (configuration II) | $1.3 \times 10^{12}$ |
| Sphere B filled with $^{90}$Y (configuration III) | $1.3 \times 10^{12}$ |
| Phantom body filled with $^{177}$Lu | $1.9 \times 10^{13}$ |
| Phantom body filled with $^{90}$Y | $2.3 \times 10^{12}$ |

For each component, the projections were scaled to the desired activity concentrations presented in table A.1 and Poisson noise was added. The activity concentrations of $^{177}$Lu in sphere A were scaled down such that when background activity was added, the total activity in the sphere would be equal to 0.1, 0.8, or 1.5 MBq/ml. The same was true for the desired activity concentration of $^{90}$Y. This allowed us to make a total of 108 activity combinations using only the original seven simulations.

Furthermore, to assess the accuracy of the results, multiple realizations for each activity combination were created. This was done by altering the Poisson noise added to each projection of each component of the simulations prior to the reconstruction. The 'new' dataset was then reconstructed and analysed in the same way as the original image. Five realizations were created for each phantom configuration, each combination of $^{177}$Lu and $^{90}$Y in the spheres, each source to background ratio and for both energy window widths. This led to a total of 540 simulations

A Siemens SymbiaT dual-head SPECT camera with a medium energy, low penetration (MELP) collimator was modelled with GATE to acquire the projection data. The camera specifications were based on the data that is publicly available from Siemens [31] and is described in detail in past works [27].

## 2.2 Simulated Projections

A total of 128 projections (64 per head) were simulated for each scan. Each scan was 16 minutes long (15s/projection) to reflect what is typically used in a clinical environment. The matrix size was 128 x 128 with a pixel size of 4.79 mm x 4.79 mm. Data was obtained in three energy windows: a lower scatter window (LW), the photopeak window centered at 208.0 keV (PW), and the upper scatter window (UW). Two different scatter windows widths were investigated, as discussed below.

## 2.3 Image Reconstruction

Image reconstructions were performed using an in-house-created graphical user interface, SPEQToR (Single Photon Emission Quantitative Tomographic Reconstruction) [32], which uses the standard ordered subset expectation maximization (OSEM) algorithm with attenuation correction, scatter correction, and resolution recovery. All reconstructions were performed with 8 subsets and 8 iterations.

To perform attenuation correction, a 128 x 128 x 128 attenuation map of a 20 cm x 20 cm cylinder (to replicate the simulated phantom) filled with linear attenuation coefficients for 208 keV photons in water (0.135 cm$^{-1}$) was created. For scatter correction, the triple energy window (TEW) method was used [24]. As Bremsstrahlung from $^{90}$Y will introduce counts into the $^{177}$Lu energy spectrum, two different scatter window widths were investigated to test how robust the TEW method performs in each scenario. A narrow window width of 3% and a wide window width of 10% were used for both the lower window (LW) and upper window (UW) (see table 3). In this study, '10%' and '3%' are always in reference to the photopeak energy window around the 208 keV. A width of 20% was always used for the photopeak window. The TEW scatter predictions were added to the estimated projections during reconstruction.

Table 3: The scatter and photopeak window widths used for triple energy window (TEW) scatter correction.

| Photopeak Window (keV) | | Scatter Window Width | Lower Scatter Window (keV) | | Upper Scatter Window (keV) | |
|---|---|---|---|---|---|---|
| Center | Range | | Center | Range | Center | Range |
| | | 3% | 184.1 | 181.0 – 187.2 | 231.9 | 228.8 – 235.0 |
| 208.0 | 187.2 - 228.8 | 10% | 176.8 | 166.4 – 187.2 | 239.2 | 228.8 – 249.6 |

## 2.4 Camera Calibration

To quantify the activity in the reconstructed images, a camera calibration factor (CF) was calculated to convert detected counts to activity values. A 16 minute (960s) planar scan of a point source containing 300 MBq of $^{177}$Lu in air was simulated. The CF calculated using a planar scan of a point source shows very good agreement with that of hot sources in a warm background (which better models patient studies but may be very cumbersome to perform) [26].

Image reconstruction is not necessary in a planar scan, however, scatter correction is required [26]. The scatter correction was performed on the projections. Like the experimental projections, two scatter window widths were used for TEW scatter correction: 3% or 10% with respect to 208 keV. The number of scatter counts in the photopeak window was calculated using the TEW method, and the number of primary photons in the photopeak window, $C_{prim}$, was determined by subtracting the scatter counts from the total number of counts in the photopeak window. The CF, in units of cps/MBq, was then determined using the equation:

$$CF = \frac{C_{prim}}{At} \tag{1}$$

where $A$ is the total activity of the source (300 MBq) and $t$ is the total acquisition time (16 minutes). The CF calculated with each of the window widths was applied to the corresponding experimental simulations.

## 2.5 Segmentation

To perform segmentation on the reconstructed images, a spherical volume-of-interest (VOI) mask was created by setting all the counts in the image matrix outside of a 3.5 cm radius sphere, centered on the true sphere location (sphere A), to zero. Although the true radius of the spheres was 3.3 cm, 0.2 cm was added to this radius to account for the spill-out of activity due to partial volume effects.

## 2.6 Data Analysis

Performing dosimetry requires calculating the time-integrated activity in an organ or tumour, which is usually derived from a series of quantitative images taken at multiple time points. Therefore, accurate activity quantification is especially important for dosimetric calculations. In our study, to assess the accuracy of activity quantification in sphere A for every simulated activity combination and configuration, the quantification error ($E$) was calculated by comparing the estimated activity to the true activity:

$$E = \frac{1}{R} \sum_{r=1}^{R} \frac{A_{est,r} - A_{true}}{A_{true}} \times 100\% \tag{2}$$

The variable $R$ is the number of realizations, which was 5 in all our simulated cases, and $A_{est,r}$ is the estimated activity in the $r$th realization. Because we were dealing with simulations, the true activity, $A_{true}$, was known exactly. Therefore, the quantification error tells us precisely how the estimated activity of $^{177}$Lu is altered due to the presence of $^{90}$Y. The error was determined separately for each configuration and ratio of $^{177}$Lu to $^{90}$Y, and TEW size.

The contrast to noise ratio (CNR) was also determined for every configuration because the presence of $^{90}$Y was expected to alter the noise in the reconstructed images, which could impact segmentation and lesion detectability. The following formula was used for CNR [33] [34]:

$$CNR = \frac{1}{R}\sum_{r=1}^{R}\frac{\left(\frac{M_{voi,r} - M_{bkg,r}}{M_{bkg,r}}\right)}{\left(\frac{\sigma_{bkg,r}}{M_{bkg,r}}\right)} \quad (3)$$

$$= \frac{1}{R}\sum_{r=1}^{R}\left(\frac{M_{voi,r} - M_{bkg,r}}{\sigma_{bkg,r}}\right)$$

where $M_{voi,r}$ is the average number of counts in the source VOI, $M_{bkg,r}$ is the average number of counts in the background VOI, and $\sigma_{bkg,r}$ is the standard deviation of the background VOI, all in the $r$th realization. The background VOI was of identical size to the source VOI and was placed inside the cylinder, below spheres A and B. The CNR was determined separately for each configuration, ratio of $^{177}$Lu to $^{90}$Y, and TEW size.

A Wilcoxon signed-rank test was used to compare the differences between both scatter window widths. A separate test was done for the CNRs and the activity quantification errors. A *p*-value less than 0.05 was considered statistically significant.

## 3. Results

### 3.1 *Camera Calibration*

The calibration factors calculated using both scatter window widths are shown in table 4.

Table 4  The calculated calibration factors

| Scatter Window Width | CF (cps/MBq) |
|---|---|
| 3% | 10.11 |
| 10% | 10.09 |

### 3.2 *Sphere Simulation*

Figures 2(a) and 2(b) show the simulated total, primary, and scattered photon spectrum originating from a sphere filled with $^{177}$Lu or $^{90}$Y, respectively, placed off-centre inside a phantom containing a radioactive solution. Figure 2(c) shows the combined $^{177}$Lu and $^{90}$Y spectrum. The spectra correspond to activities of $^{177}$Lu and $^{90}$Y in sphere A both equal to 16.6 MBq with 46.2 MBq each of $^{177}$Lu and $^{90}$Y in the phantom body.

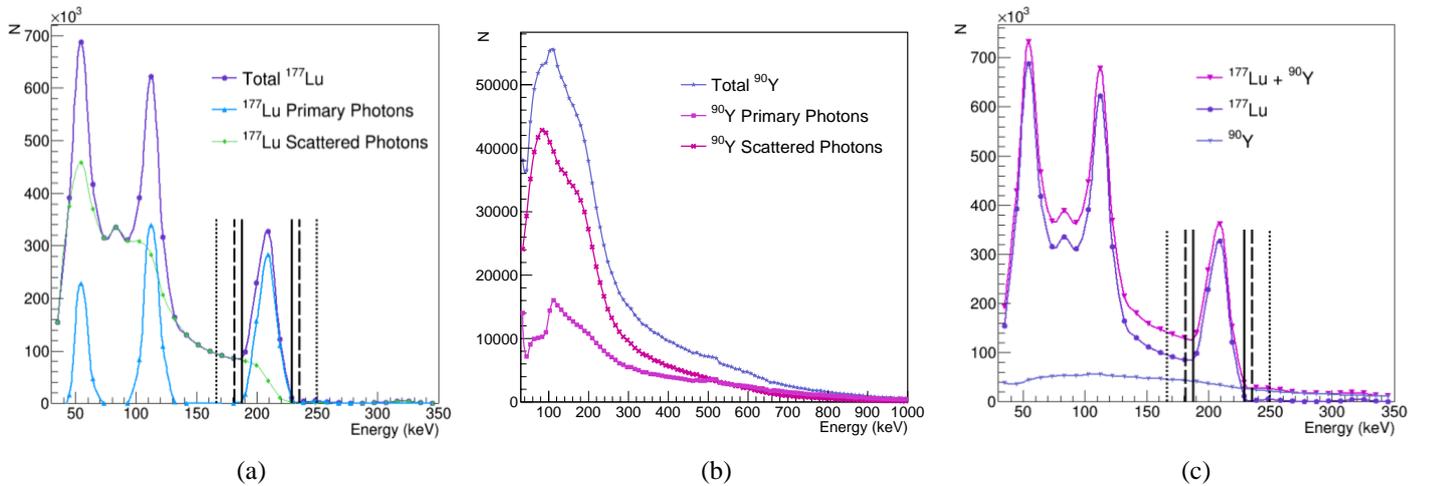

(a)　　　　　　　　　　　(b)　　　　　　　　　　　(c)

Figure 2  Energy spectra simulated with GATE as would be acquired by the SPECT camera. The dashed and dotted lines show the location of the 3% and 10% scatter windows, respectively. The solid lines show the location of the photopeak window. The activities of $^{177}$Lu and $^{90}$Y were both 16.6 MBq and there was 46.2 MBq of $^{177}$Lu and 46.2 MBq of $^{90}$Y in the phantom body. Figure 2 (a) shows the $^{177}$Lu spectra, (b) shows the $^{90}$Y spectra, and (c) shows the combined spectra.

Figure 3 shows reconstructed images for the three activity concentrations of $^{177}$Lu that were simulated in configuration II.

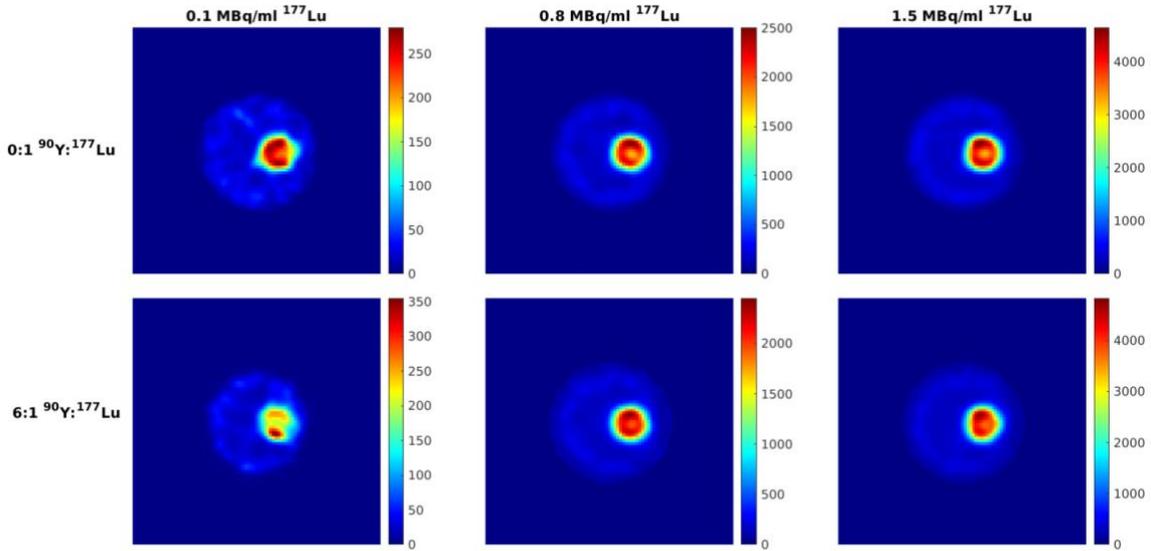

Figure 3  Reconstructed images from six of the activity combinations used in configuration II. All images have a 10:1 sphere to background ratio.

Figure 4 shows the activity quantification error and CNRs for configurations I and II. Both configurations are shown in the same plot, with configuration I containing no $^{90}$Y (Ratio of $^{90}$Y:$^{177}$Lu = 0). Figure 5 shows the same plots for configuration III.

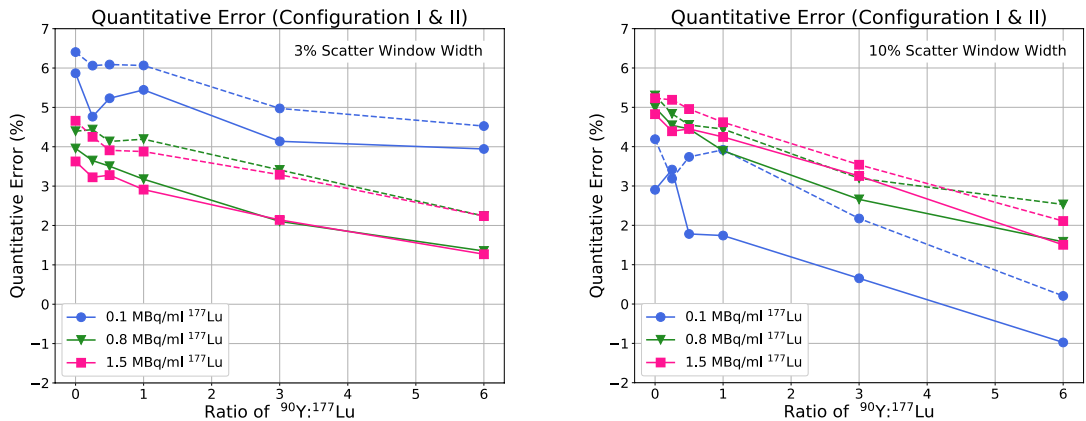

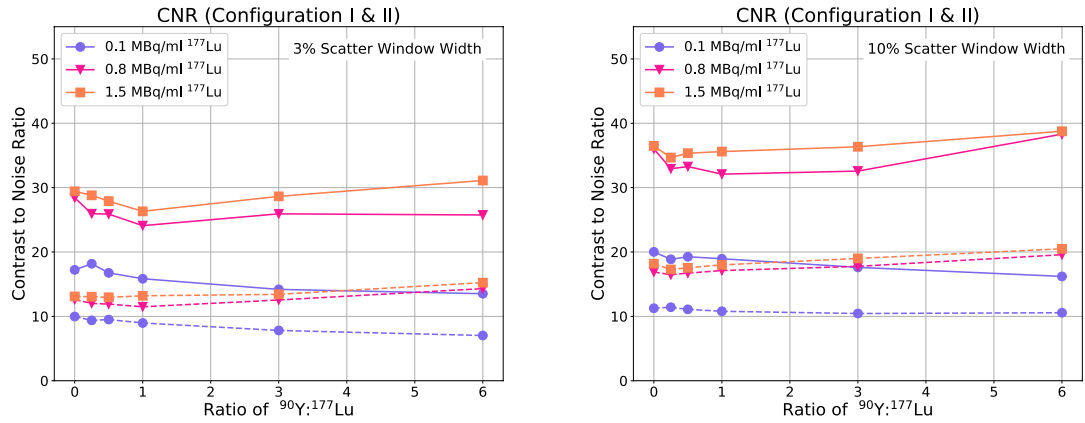

Figure 4  Activity quantification error (top) and contrast to noise ratios (bottom) for configurations I and II. The dashed lines show a 5:1 sphere to background ratio and the solid lines show a 10:1 sphere to background ratio.

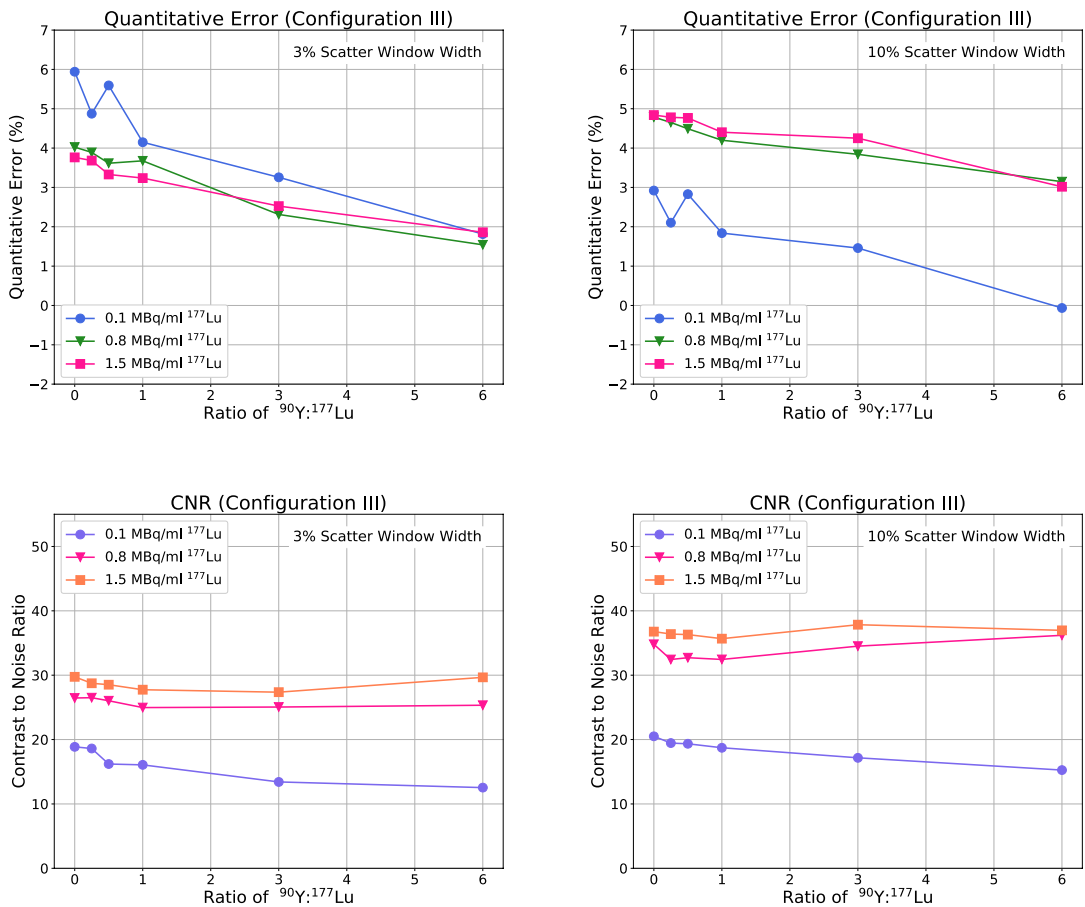

Figure 5  Activity quantification error (top) and contrast to noise ratios (bottom) for configuration III, in which $^{177}$Lu was placed in sphere A and $^{90}$Y was placed in sphere B. The solid lines show a 10:1 sphere to background ratio.

*P*-values of < 0.0001 were recovered from both of the Wilcoxon signed-rank tests. This indicates that there are statistically significant differences in the CNRs and the activity quantification errors when using the narrow versus the wide scatter window widths.

## 4. Discussion

In this study, we used simulations to determine the accuracy of quantitative SPECT imaging of $^{177}$Lu in the presence of $^{90}$Y, which needs to be sufficiently high if dosimetry in a dual-isotope RPT using $^{177}$Lu is to be performed. We also aimed to determine if large or narrow scatter window widths to use with the triple energy window scatter correction method yielded more accurate results. Finally, we aimed to determine the extent that the presence of $^{90}$Y impacts image noise by determining the CNRs of the reconstructed images.

Our study is unique as there have been no other simulation studies, to the best of our knowledge, performing SPECT imaging of $^{177}$Lu and $^{90}$Y combined. Dual-isotope SPECT imaging based on Monte Carlo simulations have been performed for radioisotope pairs such as $^{99m}$Tc/$^{123}$I [35] $^{123}$I/$^{125}$I [36], and combinations of $^{99m}$Tc, $^{111}$In, $^{123}$I, $^{177}$Lu, and $^{201}$Tl [37], but not $^{177}$Lu and $^{90}$Y. Dosimetry simulations of $^{177}$Lu and $^{90}$Y therapy have been performed [38][39] which highlights the importance of a study such as ours; dosimetry is impractical outside of a simulation if imaging cannot be performed quantitatively. Imaging has been performed in dual-isotope PRRT with $^{177}$Lu and $^{90}$Y [40], but dosimetry was not performed. We also simulated a wide range of $^{177}$Lu and $^{90}$Y source and background activity concentrations, which allowed us to generate data that is more representative of a true patient population. We analysed a total of 540 simulations, which increased the statistical power of our results.

### 4.1 *Activity Quantification*

Our results show a reduction of recovered $^{177}$Lu activity when higher $^{90}$Y to $^{177}$Lu ratios were used (see figures 4 and 5). The presence of $^{90}$Y resulted in *lower* relative quantification error compared to when only $^{177}$Lu was present. This is the opposite of what may at first be expected. A possible explanation for this is that the presence of $^{90}$Y introduced additional counts into the projections. It may be argued that this should cause the uncertainty in the recovered activity to increase, and the quantification error should become larger. However, the opposite may occur due to the additional counts reducing inaccuracies introduced by the triple energy window scatter correction (TEW SC) method. When only $^{177}$Lu was present in our phantom, TEW SC *underestimated* the total number of non-primary photons in the VOI and *overestimated* the total number of non-primary photons in the background region (outside of the VOI). When the entire image was considered, the TEW method overestimated the number of non-primary photons slightly, by 1-2%. However, this number may depend on the shape and size of the object (phantom) and the medium which is present in it. This phenomenon was confirmed in a previous study [26] and also discussed in other studies [41]. An underestimation of non-primary photons in the photopeak window leads to an overestimation of recovered $^{177}$Lu activity.

While TEW SC underestimated the total number of non-primary photons in the VOI and overestimated the total number in the background region for $^{177}$Lu, we noticed the opposite effect for $^{90}$Y. Because the Bremsstrahlung spectrum created by $^{90}$Y has no photopeaks, TEW SC should remove all the Bremsstrahlung photons from the 208 keV photopeak window of $^{177}$Lu. When we performed TEW SC on a phantom containing only $^{90}$Y (10:1 sphere to background ratio), the number of photons within the 208 keV peak window was overestimated in the VOI and underestimated in the background. This is the opposite of what was observed in the $^{177}$Lu phantom. When both isotopes are present, the underestimation or overestimation of non-primary photons by the TEW method competes. In all our simulated cases, we observed that the estimated number of non-primary photons from $^{177}$Lu in the 208 keV photopeak window became closer to the truth when more $^{90}$Y was added.

This overestimation and underestimation can be seen in figure 6, which shows a comparison between the 'predicted scattered counts' in the VOI by the TEW method and the 'true scattered counts' in the VOI (which was a 3.5 cm radius sphere centered on the true sphere, as discussed above) for different ratios of $^{90}$Y to $^{177}$Lu activity. We refer here to 'true scattered counts' as all the counts from $^{90}$Y and the non-primary counts from $^{177}$Lu that are present in the photopeak window. Non-primary counts are any photons detected by the SPECT camera that previously scattered in the phantom, collimator, or any other camera component before being detected. The 'predicted scattered counts' are the scattered counts in the photopeak window based on the TEW SC method calculation. A similar plot showing the comparison between the predicted scattered counts and the true scattered counts in the background VOI region can be found in figure A.1 in the supplementary material.

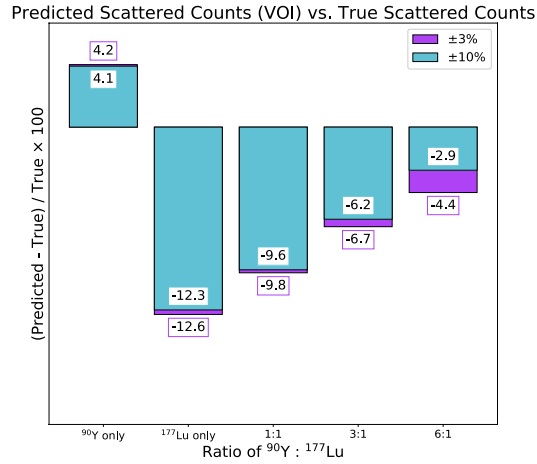

Figure 6. The difference between the 'predicted scattered counts' in the VOI by the TEW method and the 'true scattered counts' in the VOI for different ratios of $^{90}$Y to $^{177}$Lu. $^{90}$Y and $^{177}$Lu were both placed in sphere A, and the background activity concentration was 10 times lower than the activity concentration in the sphere. The activity concentration of $^{177}$Lu in the sphere was 0.1 MBq/ml.

The TEW method therefore accounts for the number of non-primary counts in the region of interest more accurately when $^{90}$Y is present. This is promising for dosimetry in dual-isotope therapies because it indicates that the presence of $^{90}$Y does not degrade activity quantification. However, the impact of this effect is minimal when using activity ratios that are used clinically. In patients, the ratio of $^{90}$Y to $^{177}$Lu is usually not much higher than 1:1 [12][10] because (as already mentioned) the higher energy emissions from $^{90}$Y are more toxic to the organs at risk than emissions from $^{177}$Lu. When the activity ratio is 1:1, the activity quantification error is within 3% of the $^{177}$Lu-only case.

The accuracy of activity quantification in sphere A for configuration III very slightly outperformed that of configuration II. In configuration II, both isotopes were in sphere A, and therefore there were more total counts from $^{90}$Y in the analysed VOI. Because of this, both primary and scattered Bremsstrahlung photons from $^{90}$Y contributed to the total activity in the VOI, while only scattered counts were contributing when $^{90}$Y was in sphere B. Regardless, the difference between configuration II and III was minimal (< 2 %).

The Wilcoxon signed-rank test showed statistically significant differences ($p$-value < 0.0001) in the activity quantification when different scatter window widths were used. At low concentrations of $^{177}$Lu (0.1 MBq/ml), the 10% scatter window width resulted in slightly lower errors, while the 3% scatter window width resulted in lower errors at higher concentrations. However, it is possible that the reduced performance of the 3% scatter window at low concentrations is due to noise induced bias. Regardless, the differences are minimal and are expected to have negligible clinical impact. Even considering the underestimation of counts in the VOI, our results suggest that quantitation of $^{177}$Lu will be within 7% for the phantom configurations used in this study, regardless of the amount of $^{90}$Y (at least up to a 6:1 ratio). This creates significant optimism for accuracy of activity quantification of $^{177}$Lu and $^{90}$Y in a dual-isotope therapy, based on $^{177}$Lu imaging only.

4.2 *Contrast to Noise Ratios*

The CNR was not impacted by the presence of $^{90}$Y, even at high $^{90}$Y/$^{177}$Lu concentration ratios. The CNR was impacted much more by the concentration of $^{177}$Lu and the amount of background activity in the cylinder. The width of the scatter windows used for TEW SC made a statistically significant difference in the CNR according to the Wilcoxon signed-rank test. The CNR was higher when the larger scatter window width of 10% was used. Regardless, the CNR remained stable and did not change much with the addition of $^{90}$Y, even though we tested a large range of $^{177}$Lu and $^{90}$Y concentrations. However, we simulated a very simple case; a patient is much larger than what we simulated here and does not have a cylindrical shape. The patient anatomy also contains different density tissues with different scattering properties. However, based on this study we would expect noise and lesion detectability to not be an issue when imaging $^{177}$Lu in the presence of $^{90}$Y, even in a patient.

Based on our results, we would recommend using a larger scatter window width for TEW scatter correction. As discussed, there is a very small difference in activity quantitation between the two different scatter window widths (1-2%) but the larger scatter window width resulted in less noise.

Finally, we would like to acknowledge some limitations in our study. Monte Carlo simulations are excellent models of physics phenomena, but they are not identical to a real phantom experiment. The quantitation accuracy that we achieved may be higher than what would occur in a real phantom study because we can closely control the simulation environment. We also did not investigate the effect of different segmentation methods on our results. The choice of segmentation method may alter the shape and size of the VOIs which may alter our results. Further studies analysing different segmentation methods on $^{177}$Lu quantitation in the presence of $^{90}$Y would be beneficial to determine the impact of segmentation. A next step to our study involves expanding simulations to more realistic patient geometry or performing a phantom study with $^{177}$Lu and $^{90}$Y. Our overarching goal is to perform accurate dosimetry leading to better patient outcomes, and these types of simulations can help us further understand and improve dosimetry for neuroendocrine tumours.

## 5. Conclusion

We simulated a phantom study of spheres with different concentrations of $^{90}$Y and $^{177}$Lu placed in a water phantom to validate the accuracy of quantitative SPECT imaging of $^{177}$Lu in the presence of $^{90}$Y. Our results suggest that image quantification remains acceptable (within 7%) even when very large concentrations of $^{90}$Y (6 times higher than the amount of $^{177}$Lu) are present. We also tested the TEW scatter correction method and saw significant differences in the activity quantification error between scatter window widths. The differences, however, were small (within 2%) and are expected to have no detrimental impact in the clinical environment. We also showed that the presence of $^{90}$Y did not impact the CNR. The CNR was impacted significantly by the concentration of $^{177}$Lu and the amount of activity in the background, but not the amount of $^{90}$Y. The CNR is higher when the larger scatter window width of 10% is used. Based on these results, we conclude that quantitative SPECT imaging of $^{177}$Lu remains accurate in the presence of $^{90}$Y, and $^{90}$Y does not negatively impact activity quantification or lesion detectability.

## Acknowledgements

The authors gratefully acknowledge funding by the Natural Sciences and Engineering Research Council of Canada (NSERC) Grants RGPIN-2017-04914, RGPIN-2019-06467, and PGSD-411297574. We also gratefully acknowledge Dr. Jean-Mathieu Beauregard (Quebec City, Quebec, Canada) for providing us with data on which we based our simulation studies.

# Supplementary Material

| Configuration | Activity concentration in 150.5 ml sphere (MBq/ml) | | Total activity in 150.5 ml sphere (MBq) | | Ratio of activity in spheres | Activity concentration in 6283 ml cylindrical phantom body (MBq/ml) | | Total activity in 6283 ml cylindrical phantom body (MBq) | |
|---|---|---|---|---|---|---|---|---|---|
| | $^{177}$Lu | $^{90}$Y | $^{177}$Lu | $^{90}$Y | $^{90}$Y:$^{177}$Lu | $^{177}$Lu | $^{90}$Y | $^{177}$Lu | $^{90}$Y |
| I | 0.1 | 0 | 15.0 | 0 | 0:1 | 0.01, 0.02 | 0 | 62.8, 125.7 | 0 |
| | 0.8 | | 120.4 | | | 0.08, 0.16 | | 502.6, 1005.3 | |
| | 1.5 | | 225.8 | | | 0.15, 0.3 | | 942.5, 1884.9 | |
| II | 0.1 | 0.025 | 15.0 | 3.8 | 0.25:1 | 0.01, 0.02 | 0.0025, 0.005 | 62.8, 125.7 | 15.7, 31.4 |
| | | 0.05 | | 7.5 | 0.5:1 | | 0.005, 0.01 | | 31.4, 62.8 |
| | | 0.1 | | 15.0 | 1:1 | | 0.01, 0.02 | | 62.8, 125.7 |
| | | 0.3 | | 45.2 | 3:1 | | 0.03, 0.06 | | 188.5, 377.0 |
| | | 0.6 | | 90.3 | 6:1 | | 0.06, 0.12 | | 377.0, 754.0 |
| | 0.8 | 0.2 | 120.4 | 30.1 | 0.25:1 | 0.08, 0.16 | 0.02, 0.04 | 502.6, 1005.3 | 125.7, 251.3 |
| | | 0.4 | | 60.2 | 0.5:1 | | 0.04, 0.08 | | 251.3, 502.6 |
| | | 0.8 | | 120.4 | 1:1 | | 0.08, 0.16 | | 502.6, 1005.3 |
| | | 2.4 | | 361.3 | 3:1 | | 0.24, 0.48 | | 1507.9, 3015.8 |
| | | 4.8 | | 722.5 | 6:1 | | 0.48, 0.96 | | 3015.8, 6031.7 |
| | 1.5 | 0.375 | 225.8 | 56.4 | 0.25:1 | 0.15, 0.3 | 0.0375, 0.075 | 942.5, 1884.9 | 235.6, 471.2 |
| | | 0.75 | | 112.9 | 0.5:1 | | 0.075, 0.15 | | 471.2, 942.5 |
| | | 1.5 | | 225.8 | 1:1 | | 0.15, 0.3 | | 942.5, 1884.9 |
| | | 4.5 | | 677.4 | 3:1 | | 0.45, 0.9 | | 2827.4, 5654.7 |
| | | 9 | | 1354.8 | 6:1 | | 0.9, 1.8 | | 5654.7, 11309.4 |
| III | 0.1 | 0.025 | 15.0 | | 0.25:1 | 0.01 | 0.0025 | 62.8 | 15.7 |
| | | 0.05 | | | 0.5:1 | | 0.005 | | 31.4 |
| | | 0.1 | | | 1:1 | | 0.01 | | 62.8 |
| | | 0.3 | | | 3:1 | | 0.03 | | 188.5 |
| | | 0.6 | | | 6:1 | | 0.06 | | 377.0 |
| | 0.8 | 0.2 | 120.4 | | 0.25:1 | 0.08 | 0.02 | 502.6 | 125.7 |
| | | 0.4 | | | 0.5:1 | | 0.04 | | 251.3 |
| | | 0.8 | | | 1:1 | | 0.08 | | 502.6 |
| | | 2.4 | | | 3:1 | | 0.24 | | 1507.9 |
| | | 4.8 | | | 6:1 | | 0.48 | | 3015.8 |
| | 1.5 | 0.375 | 225.8 | | 0.25:1 | 0.15 | 0.0375 | 942.5 | 235.6 |
| | | 0.75 | | | 0.5:1 | | 0.075 | | 471.2 |

| | | | | | | | |
|---|---|---|---|---|---|---|---|
| | | 1.5 | | 1:1 | | 0.15 | 942.5 |
| | | 4.5 | | 3:1 | | 0.45 | 2827.4 |
| | | 9 | | 6:1 | | 0.9 | 5654.7 |

Table A.1  Summary of the activity concentrations of $^{177}$Lu and $^{90}$Y in the spheres and phantom body for each configuration.

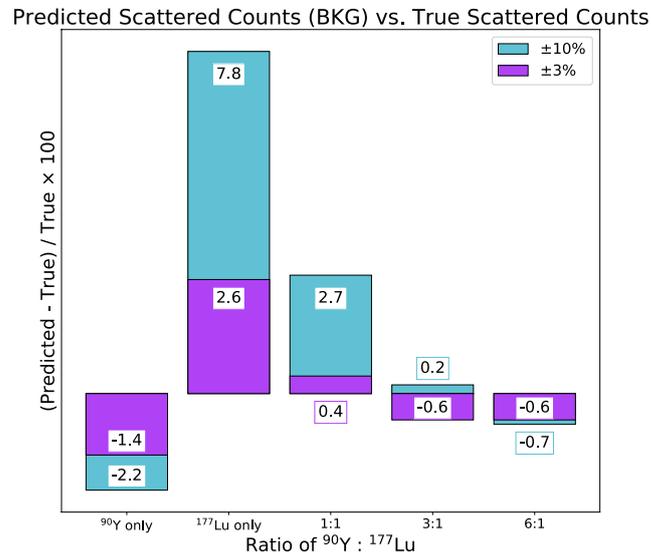

Figure A.1  The difference between the 'predicted scattered counts' outside of the VOI by the TEW method and the 'true scattered counts' outside of the VOI for different ratios of $^{90}$Y to $^{177}$Lu. $^{90}$Y and $^{177}$Lu were both placed in the same sphere, and the background activity concentration was 10x lower than the activity concentration in the sphere.